\documentclass[manuscript]{aastex}

\shorttitle{Eclipsing Binaries in Young Globular-Like Cluster NGC 1850}
\shortauthors{Taylor}

\begin{document}

\title{Eclipsing Binaries in the Young LMC Cluster NGC 1850}


\author{Stuart F. Taylor}
\affil{Center for High Angular Resolution Astronomy (CHARA),
 Physics and Astronomy Department, Georgia State University,
    Atlanta, GA 30303}
\email{taylor@chara.gsu.edu}

\begin{abstract}


I present light curves for two detached eclipsing binary
stars in the region of the LMC cluster NGC 1850, 
which is possibly a young globular cluster still in formation.  
One, a likely spectral type O star,
is a newly detected eclipsing binary in the
region of the very young subcluster NGC 1850A.  
This binary is among a small number 
of highly massive O-type stars in binary systems found in LMC clusters. 
These two eclipsing binaries are the first 
discovered in the well studied NGC 1850,
and the O-type star is the first eclisping binary found in NGC 1850A.
Light curves for two NGC 1850 region Cepheid variables are also 
shown.  Discovering two eclipsing binaries in the 
young globlular-like cluster NGC 1850 
is discussed in terms of the importance of the binary fraction to 
globular cluster evolution.  

\end{abstract}



\section{Introduction}



Binary stars influence globular cluster evolution to a far greater 
degree than had long been believed \citep{mey97,hut92}.
Once thought not to exist in numbers significant 
enough to influence the 
evolution of globular clusters (GCs), binary stars are now 
seen as providing energy to GC single stars and 
halting core collapse.   
Not only has the importance of binary stars increased to 
theoretical studies, observers are now detecting globular 
cluster binaries, overturning earlier observational reports 
that GCs do not contain signficant numbers of primoridial 
binary stars \citep{mey97}.

The question that arises is whether the binary fraction in 
GCs changes with age.  While young globular 
clusters have been studied theoretically, it is well known 
that Milky Way GCs were all formed in the first 
few Gyr of the history of the universe, but such is apparently 
not the case in the Magellanic Clouds, with LMC globular-like clusters (GLCs,
also called ``young, globular-like stellar associations'')
such as NGC 1850 and NGC 1818, which have ages on the orders of $10^7$ or 
even $10^6$ years.  Preliminary observational work has led to the 
discovery of some binaries in the field near these clusters \citep{seb95},
but surveys are needed in order to compare the binary fraction 
of these ``young'' GLCs with that of the better 
studied but older Milky Way GCs.

Observations are needed to address the central question of 
how binaries affect the dynamical evolution 
of {\it young} globular-like clusters.
The detection reported here of two eclipsing binary stars 
in the region of the young GLC NGC 1850 does not
yet answer this question, but shows that the use of moderate sized
(0.9 m) telescopes can be brought to bear on constraining the answer.  


I present new time series imaging of the LMC cluster NGC 1850 and an
analysis of photometric variables using image subtraction methods.
I present four light curves for NGC 1850 region systems.
First, I discuss the finding of a second detached eclipsing
binary in the region of NGC 1850 (refered to here as ``Variable 5'' or ``V5'');
which is the first eclipsing binary found in the region of
the subcluster NGC 1850A.  Next, I present an eclipsing binary 
(``Variable 1'' or ``V1'')
in the larger region of NGC 1850 which was found by 
the Optical Gravitational Lens Experiment (OGLE) collaboration 
\citep{uda03}, and its light curve is presented here.  
The OGLE group has graciously
confirmed that their data show V5 to be an eclipsing binary that
was originally missed in their large, comprehensive catalog of MC
eclipsing binaries 
\citep{uda99,uda03} 
(publicly available from the OGLE project 
website\footnote{http://sirius.astrouw.edu.pl/\textasciitilde
ogle/ }),
and data from the much longer-term 
OGLE project provided a better determination of the
period than the current dataset alone (A. Udalski, private communication).

\section{Observations and Pre-Processing}

Seven nights of observations of a field roughly centered on NGC 1850
were taken during 2002 February 20-23 and March 03-05, with the 0.9 m 
Cerro Tololo Inter-American Observatory (CTIO) telescope.  
These observations are summarized in Table \ref{obsTable}, 
and where taken in the ``Cousins $I$ ($I_C$)'' band.
This telescope was equipped with a 
2048 $\times$ 2048 Tektronix CCD camera with 0\farcs401 
pixel$^{-1}$ plate scale (as measured by Jao et al. 2003).

Bias frames and dome flats were taken at the beginning of 
each night.  All frames were read through four amplifilers, one
for each quarter of the frame.
The raw data were reduced using the standard
Image Reduction and Analysis Facility 
(IRAF\footnote{IRAF is distributed by the National Optical Astronomy
Observatory, which is operated by the Association of Universities 
for Research in Astronomy, Inc., under cooperative agreement 
with the National Science Foundation.}) 
tasks ZEROCOMBINE and FLATCOMBINE.  
The task XCCDPROC was used to combine the four subframes produced
by the four amplifiers. 

\section{Data Reduction and Analysis: 
Image-Subtraction and Light Curves}

Relative photometric light curves for 
photometric binary stars were extracted using 
the image-subtraction package 
ISIS\footnote{Available at http://www2.iap.fr/users/alard/package.html} 
\citep{ala98,ala99,ala00},
from the {\it I}-band data.  Image-subtraction was chosen to
deal with the highly crowded fields.  Image-subtraction 
is more effective than traditional photometry for detection of 
variable stars \citep{bon03}, and it works especially well in highly 
crowded fields \citep{woz00,zeb01}.  


Use of the ``optimal image-subtraction'' method of 
Alard (1999) and Alard \& Lupton (1998) with Alard's 
ISIS image-subtraction reduction packages requires going 
through the following steps: 

1) Image alignment: A new version of each image is 
produced that is aligned to one of the images that 
has been chosen as the grid ``register'' image.  The 
alignment is done by matching locations of stars, 
and can actually work better for a moderately crowded 
field.  The matching is done to a fraction of a pixel 
by interpolation, with the new images not only shifted 
and rotated but also adjusted 
for spatial scale variation within the image.

2) Reference image creation: From a user-selected 
list of the best seeing images, one combined reference image 
is created by stacking the best images but removing 
inconsistent signals that are likely to be cosmic rays 
and other defects.

\begin{figure}
\epsscale{.45}
\plotone{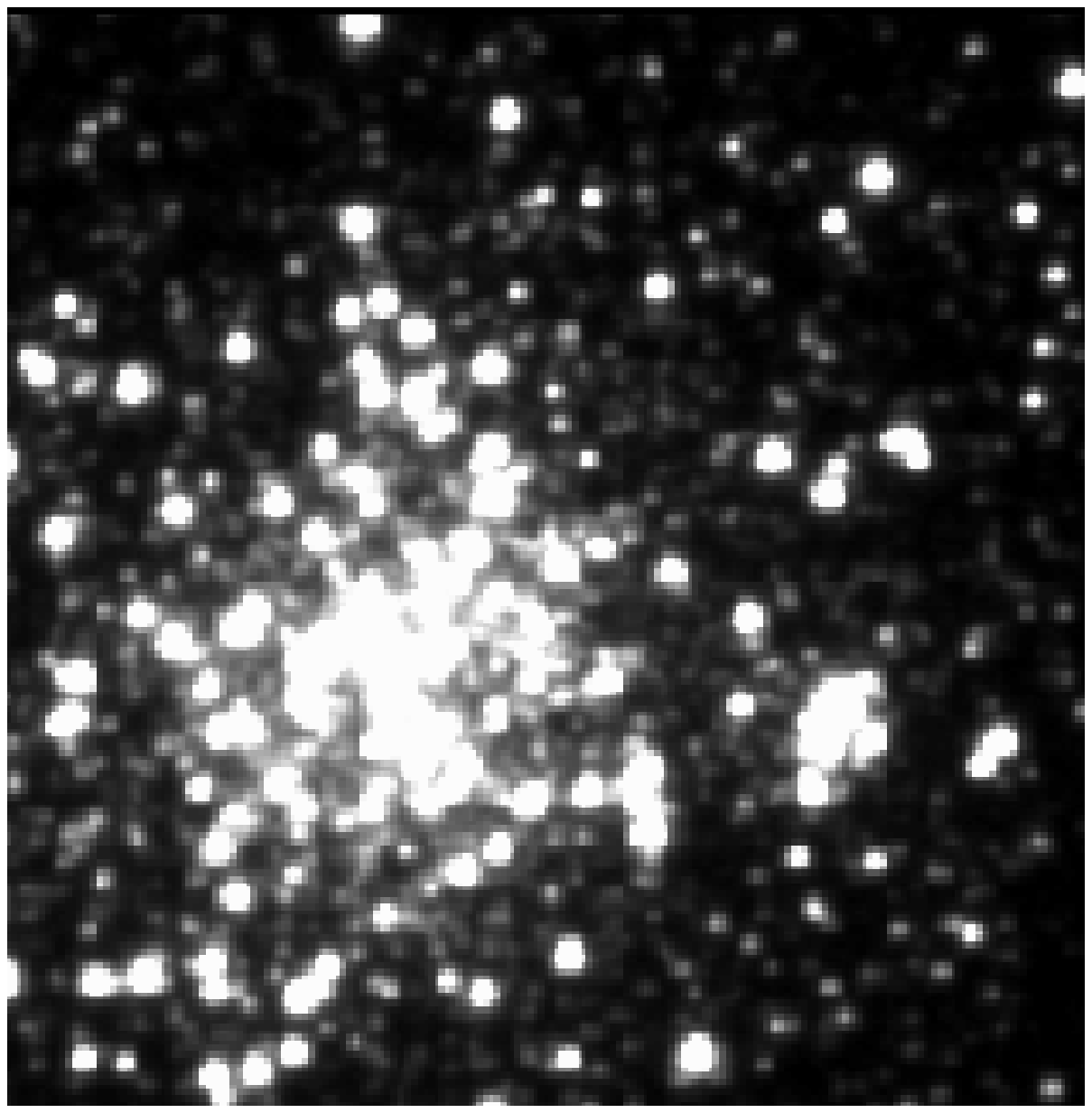}
\vspace{2mm} \hspace{1mm}
\plotone{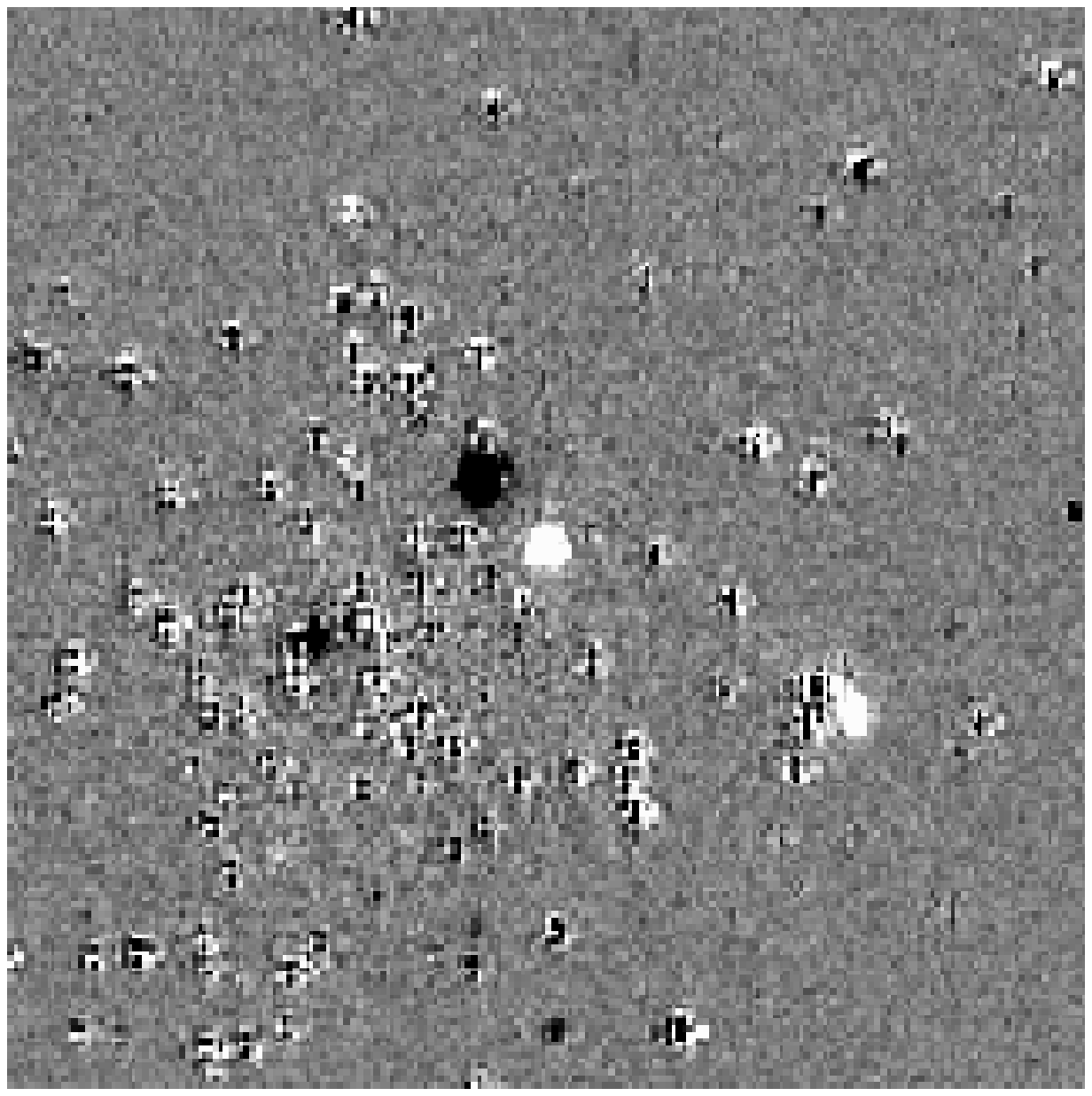}
\caption{ ($a$) The ``reference'' image of NGC 1850 (left) 
along with ($b$) the same region from an image subtraction frame (right).
These images show the central region of NGC 1850 
in an 80 arcsecond square frame (north up, east left).  The central region
of the main body of NGC 1850 is seen below left of center, and NGC 1850A 
is the small cluster of stars to its right.
The subtracted image is the difference between the reference and 
individual images.  
Stars such as V5 and V2 that were brighter during the time of the reference 
image show as white PSFs, while stars such as SW 17 and V4 which were brighter 
during the time of this particular frame 
appear as black PSFs.  V2 and 
SW 17 stand out as a pair near the center of the image with strong 
residual PSFs (SW 17 is the prominent black PSF, and V2 is the prominent
white PSF).  V5 is the prominent white PSF on the right side of NGC 1850A.
V4 is also apparent as a black PSF slightly left of 
the main NGC 1850 body's center.
V1 was not in eclipse during either time, and thus leaves no PSF in this
subtracted image.}
\label{cmpImgSub}  
\end{figure}

3) Formation of an optimum convolution kernel and image subtraction: 
For each image, a PSF matching ``kernel'' image is produced that 
minimizes least squares differences between each image and the 
reference image.  This is a big improvement over previous 
methods because it is not necessary to know either the PSF 
or the backgrounds of individual images.  Also, this only requires 
artificially degrading the seeing of the best images to that 
of each individual image, rather than as in earlier image subtraction 
efforts where all images were degraded to the seeing of the 
worst image.  ISIS had difficulty processing the full 
13\farcm6 $\times$ 13\farcm6 frames, but ran successfully
when the frames were cropped to 12\farcm0 $\times$ 12\farcm0.
On the subtracted images, the constant stars ideally will
cancel out, with only a signal from the variable stars
remaining (Fig. \ref{cmpImgSub}); in practice, 
residual signals from brighter stars
remain, complicating identification of varying stars.
Fortunately, true variable stars have smooth difference
signals with PSFs resembling that of stars, 
while imperfections in the PSFs of bright stars create residuals 
with a characteristic wave pattern
(as can be seen in Figure \ref{cmpImgSub}). 
This generally allows truly varying stars to be identified,
but the high amount of residual noise in the densely
crowded region towards the center
of the cluster may obscure variable stars that might
otherwise be identifiable.
The presence of noise from brighter stars 
is a consequence of their 
larger noise variation from frame to frame than the
background; hence, while the difference between PSFs of faint 
stars are low enough that the fainter stars' differenced
signals fade into the background, the brighter stars'
differenced signals still stand above the background.

4) Variable star identification: From the subtracted images, 
images of the normalized mean absolute deviation 
are stacked to create a variable star finding image, 
in which variable stars stand out well, 
but defects from a single image also show up.  
The ISIS software identifies variations that have more than a single 
increase in an individual image to accept a PSF-fitted 
region as a candidate variable star, thus rejecting the 
majority of cosmic rays.  The routine does not do as well 
at rejecting variations from bad columns, but these are easily recognizable  
by eye and edited out by hand.

5) Light curve photometry:  Differenced photometric 
light curves as a function of time are produced from the image 
subtracted images, using both conventional photometric methods 
of PSF-fitting and aperture photometry.  A list file of these light 
curves giving positions and other properties of the light curves 
is also generated.  Because these light curves are 
created from image subtracted images rather than the original, 
the differential signal is output, and because the images are 
relative to a reference image, the values are negative as 
well as positive.  While the disadvantage 
of image subtraction is that the 
magnitude of each star is left to be determined after the image 
subtraction analysis is completed, the advantage is that smaller 
relative differences from image to image stand out more strongly.

6) Period Finding and light curve fitting: The algorithm of 
Schwarzenberg-Czerny \citep{sch96},
optimized for searching for a periodicity in unevenly sampled observations 
is used to obtain preliminary values for the periods,
though the periods must be plotted and checked by eye.
Subsequently, the Analysis of Variance code 
written and provided by A. Udalski \citep{uda03}
was found to yield better-fitting periods (both as measured 
against previously found periods and as measured by eye).

For each variable star light curve presented here, 
the differenced flux counts $F_{difference}$ are 
divided by the normalized flux counts from 
the reference image $F_{total}$, using PSF profile-fitting photometry. 
(ISIS outputs aperture photometry counts as well.)  
The magnitude difference 
$\Delta m$ is the usual $2.5 \log{(F_{difference}/F_{total})}$.  

A detailed description of using ISIS in practice is given by 
Bruntt\footnote{Available at 
http://astro.phys.au.dk/\textasciitilde
bruntt/tuc47.html} (2003). 


\section{Photometric Variable Stars in NGC 1850 Field of View Region}


Here I present relative photometric light curves for four 
stars in the NGC 1850 region, including two cluster detached eclipsing
binary (EB) stars (V1 and V5) and two cluster Cepheid variables
(V2 and V4); data for 
a nearby LMC field star is presented in the next section.  
These variables are summarized in Tables 
\ref{varsTable}, \ref{oglePhotTable}, \ref{varsIDtable}, and 
\ref{positionsTable}, which extensively use data from the OGLE 
project \citep{uda99,uda03} obtained online and through the 
gracious private communication of A. Udalski (2003).  Both OGLE II 
and OGLE III data were used, OGLE II data for its more rigorous  
absolute calibration, and the still preliminary OGLE III data 
for its better relative precision. 
Table \ref{varsTable} gives the periods of the variables to within
roughly 0.01 day, 
and the amplitudes of the variation with errors of less than 0.1 
magnitude.  For the eclipsing binaries, the two values are for 
the primary and secondary eclipses.  
Table \ref{oglePhotTable} gives the variables' OGLE II photometry 
in $V$, $B$, and $I$ magnitudes. 

\begin{figure}
\epsscale{.50}
 \resizebox*{0.45\textwidth}{!}{\includegraphics{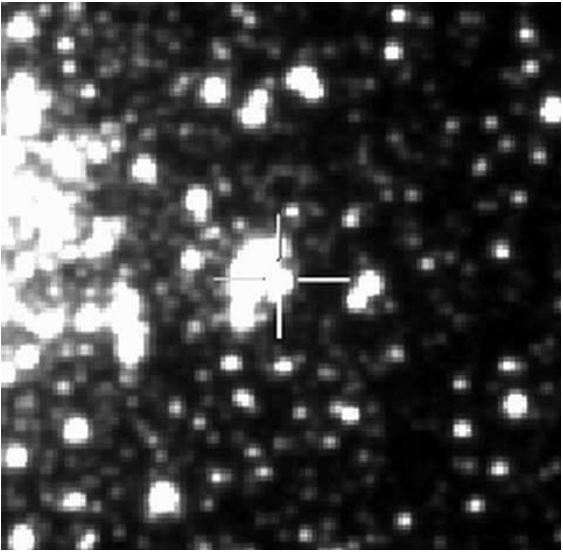}}
\caption{ Finder chart for V5 in a 60 arcsec square field, 
with North up and East left.}
\label{v6-find}
\end{figure}

\begin{figure}
  \epsscale{.50}
  \rotatebox{90}{\plotone{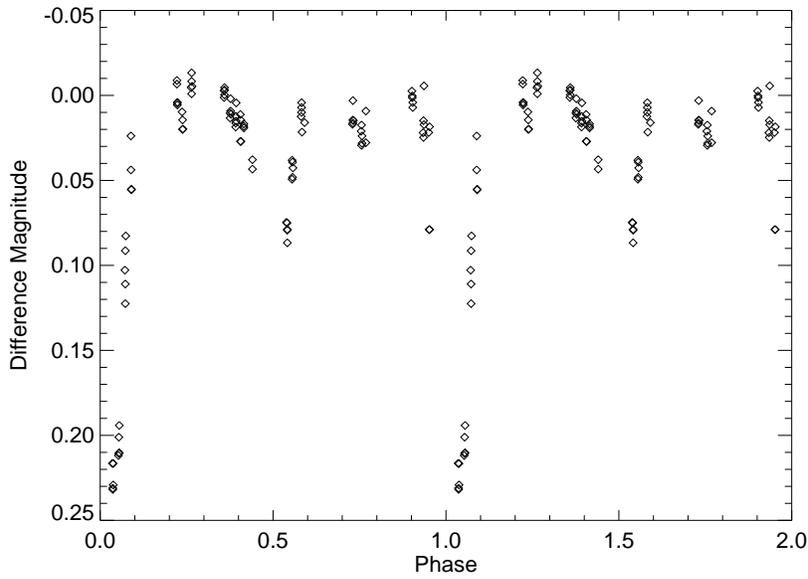}}
  \caption{ 
  $I$-band difference magnitude for phased light curve of V5, a newly-found 
    detached eclipsing binary in the central region 
    of the young subcluster of NGC 1850 referred to as NGC 1850A. 
  }
  \label{stv006-dmag}
\end{figure}

The most significant find is V5, 
a detached eclipsing binary with a period of 3.13 days that, 
unless it is a very coincidentally positioned LMC foreground star, 
is probably a member of the young subcluster NGC 1850A. 
The subcluster has an age in
the range of 4 Myr to 6 Myr as 
reported by \citet{seb95} and \citet{gil94}, 
although \citet{cal98} report an age spread of about 10 Myr.  However, 
Caloi \& Cassatella point out that if massive
cluster stars are often members of binary systems, then this
larger age spread may not be valid.
The position of V5 is shown in Figure \ref{v6-find}, and 
the light curve of V5 is shown in Figure \ref{stv006-dmag}.

V5 is in a group of a small number of young high luminosity stars that make
NGC 1850A's core stand out brightly from the rest of NGC 1850.  
The strong blending between V5 and the other bright subcluster stars
is the likely reason its binary nature was missed by previous surveys.

\begin{figure}
\epsscale{.50}
\resizebox*{0.45\textwidth}{!}{\includegraphics{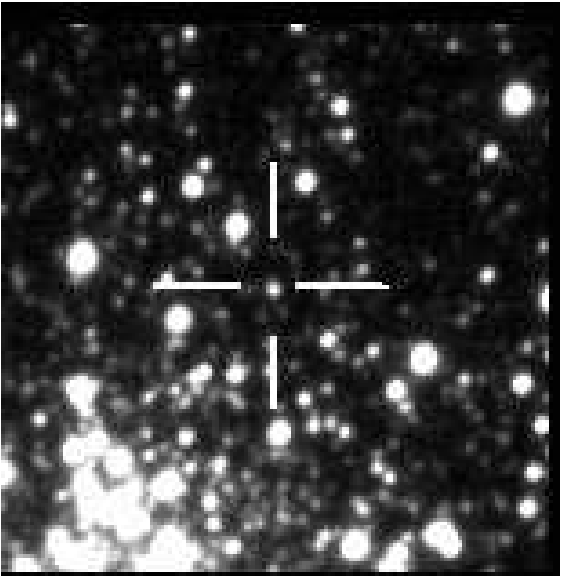}}
\caption{ Finder Chart for V1, a detached eclipsing binary 
in the region of NGC 1850,
in a 60 arcsec square field, with North up
and East left.}
\label{nov03ebs-nuel}
\end{figure}

\begin{figure}
  \epsscale{.450}
  \rotatebox{90}{\plotone{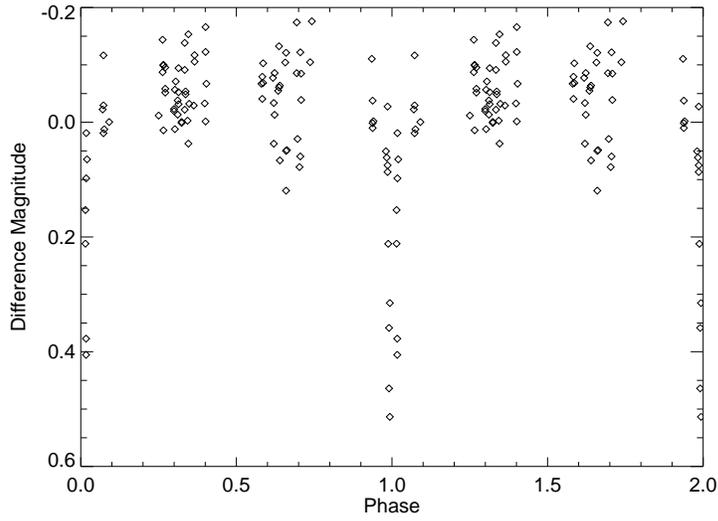}}
  \caption{ $I$-band difference magnitude phased light curve of V1, a
    detached eclipsing binary likely in the region of NGC 1850.
  }
  \label{stv001-dmag}
\end{figure}

The second eclipsing binary in the cluster region, V1, was previously 
noted in the OGLE collaboration's LMC eclipsing 
binary stars catalog \citep{uda03}, and V1 is also a detached 
eclipsing binary with a 1.48 day period.
Its position is shown in Figure \ref{nov03ebs-nuel}, and 
its light curve is shown in Figure \ref{stv001-dmag}.  
Though within the cluster radius, the NGC 1850 region's 
location in the fairly crowded LMC field will require confirmation
of V1's cluster membership.

The $B$, $V$, and $I$ magnitudes give 
partial confirmation that V5 and V1 are members of the NGC 1850 group, 
because as will be shown their luminosities are consistent 
with the LMC mean distance modulus of $18.50 \pm 0.13$ \citep{pan91},
though they still could be LMC foreground stars.  I use the reddening 
value for NGC 1850 from \citet{gil94} of $E(B-V)$ = $0.18 \pm 0.02$ mag, 
which gives an interstellar extinction of 
$A_V = 3.1 E(B-V) = 0.56 \pm 0.06$ mag.             
Using the OGLE II photometry for V5 of $m_V = 14.40 \pm 0.12$ mag, 
this interstellar extinction and distance modulus  
gives an absolute magnitude of $M_V = -4.66 \pm 0.19$. 
For a single star, $M_V = -4.7$ corresponds to V5 being of 
spectral type O5 or O6 \citep{han97}.  Being binary, V5 could have 
a primary between $M_V = -4.7$ to -3.9 mag, going 
from the case of a dim secondary to the case of 
a pair of equal luminosity type O8 stars. 
Spectral types O5/O6 have $(B-V)$ = -0.30 mag and spectral type 
O8 has $(B-V)$ = -0.285 mag \citep{weg94}.  
OGLE II photometry gives the 
$B$ magnitude of V5 as $ 14.23 \pm 0.10$          
which gives a raw $(B-V) = -0.17 \pm 0.15$ mag.   
When adjusted for redenning 
V5 has a resulting measured 
$(B-V)$ value of $-0.35 \pm 0.15$ mag, which is only slightly higher 
than $(B-V) = -0.30$ mag expected from an O5 magnitude 
star at the NGC 1850's reddening value.  
(Other authors' LMC distance moduli vary more than this 
$\pm 0.12$ mag error, and $(B-V)$ values are not definitive for O stars, 
so V5 could be an earlier or later type O star, which would not change 
the distance conclusion here.) 
While the magnitudes are consistent with V5 being an NGC 1850 member, 
the difference between NGC 1850's (and NGC 1850A's) and the LMC's 
distance modulus is not well 
known \citet{gil94}, so the measured values would also be consistent with V5 
being an LMC background star.  
Since only its LMC membership is established, confirmation of 
the membership of V5 in NGC 1850A is important to using V5 to 
study NGC 1850A.

Similarly, the OGLE photometry of V1, $m_V = 18.09 \pm 0.17$ mag
and the LMC distance modulus and the NGC 1850 interstellar reddening used 
above give V1 an $M_V = -0.97 \pm 0.22$ mag corresponding 
to a type B2 to B3 star (dim secondary) which have 
$(B-V) = $ -0.23 to -0.18 mag ranging to a pair of $M_V = -0.2$ mag type B5 
stars which have $(B-V) = -0.15$ mag \citep{weg94}. 
The OGLE  photometry for V1 of $B$ magnitude of $18.16 \pm 0.20$ which 
gives a raw $(B-V) = 0.07 \pm 0.26$ mag that when adjusted for 
reddening is $(B-V) = -0.11 \pm 0.27$ mag.  This is in good agreement with the 
the LMC distance modulus, though V1 will require further study to exclude 
the possibility of its being an LMC foreground star.

Definitive confirmation of V5 and V1 as cluster members would put 
the currently known NGC 1850 binary star census at two (both detached).
It would be worthwhile to obtain spectroscopic orbits which combined 
with light curves would 
provide direct distances to the NGC 1850 and NGC 1850A \citep{and91,fit03}.
Determination of separate distances to V1 and V5 
has the potential of providing 
distance measurement between the main body and NGC 1850A, as well as 
a better measurement of NGC 1850's position within the LMC. 
This is especially important as ``double'' or ``multiple'' clusters 
in the LMC have a significant chance of being merely chance 
superposition \citep{die02}.

Spectra to provide an accurate spectral type for V5 would 
be of special interest as it is  
one of only a relatively small number of early to middle type O 
binaries known.  A recent catalog of spectroscopic binary 
orbits\footnote{Available at http://www.chara.gsu.edu/\textasciitilde
taylor/catalogpub/ } published up to 2001 June, 
``The CHARA Spectroscopic Binary Catalog'' \citep{tay03}, is available, 
which lists 
70 type O star orbits among its total of 2353 orbits, with 12 of 
these stars being type O6 and earlier, nine of type O7, 19 O8, and 25 O9.
In 2002, new orbits were published by \citet{mas02} bringing an
additional three orbits for stars of massive type O3 and one of type O5, 
but the number of young detached stars with known masses at the highest 
stellar masses remains small \citep{gie03}.


In addition to the early-type binaries described above, 
several previously known Cepheid variables were also clearly seen
in the image subtracted current dataset.  However, the 
phase coverage was only sufficient to produce two clear Cepheid light
curves, labeled here as V2 and V4.

\begin{figure}
\epsscale{.50}
\resizebox*{0.45\textwidth}{!}{\includegraphics{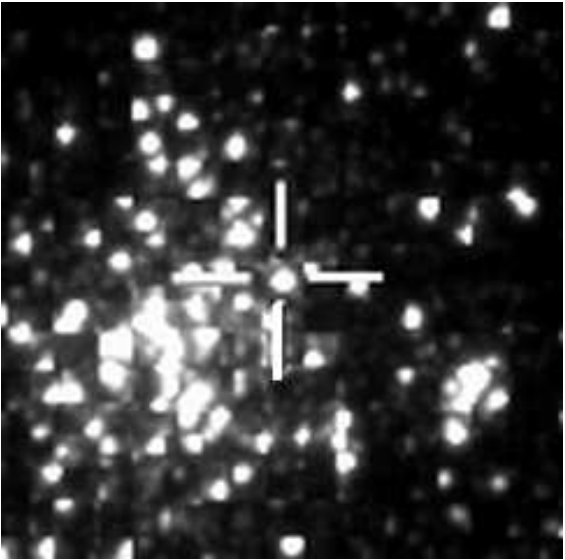}}
\caption{ Finder chart for V2 
(OGLE Cepheid LMC\_SC11 250938) in a 60 arcsec square field, with North up
and East left. }
\label{feb04cep-find}
\end{figure}

\begin{figure}
  \epsscale{.450}
  \rotatebox{90}{\plotone{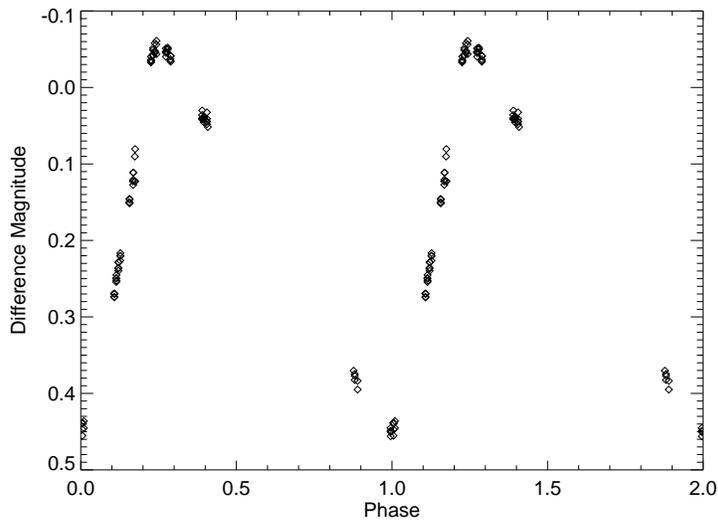}}
  \caption{ $I$-band difference magnitude phased light curve of V2, a 
    Cepheid variable that appears in the line-of-sight region of
    NGC 1850, but with a period-derived luminosity consistent 
    with the distance of the cluster (Sebo \& Wood 1995).
    }
  \label{stv002-dmag}
\end{figure}

V2 is a cluster Cepheid variable with an 8.56 day period, 
was known to Sebo \& Wood and earlier authors \citep{seb95},
and is listed as SW 58.  It is identified in the 
OGLE internet catalog of LMC Cepheid variables \citep{uda99} 
and is in the MACHO LMC survey internet 
database\footnote{http://wwwmacho.mcmaster.ca/}.
The position of V2 is shown in Figure \ref{feb04cep-find}, and
its light curve is shown in Figure \ref{stv002-dmag}.  
The difficulty of measuring the magnitude of a star in such 
a crowded core region of 
stars is illustrated
by the MACHO data where only 
141 of the 1215 $B$ band and 120 of the 530 $R$ band measurements of V2
had errors low enough to use the measurements.  
V1 and V4 are in similarly difficult to measure crowded regions.

V2 shows brightly in many image subtracted frames such as
Figure 1b, 
where it is the southwest of the two 
strongest image subtracted signals in the frame.  
The other prominent nearby signal is ``SW 17'' (Sebo \& Wood Variable 58), 
which has a period longer than covered by the current data.  
Despite being in the cluster core region, 
Sebo \& Wood found SW 17's period-derived luminosity to be inconsistent with 
other NGC 1850 Cepheids, particularly V2 (SW 58), indicating that 
SW 17 is a foreground star \citep{seb95}.  

\begin{figure}
\epsscale{.50}
\resizebox*{0.45\textwidth}{!}{\includegraphics{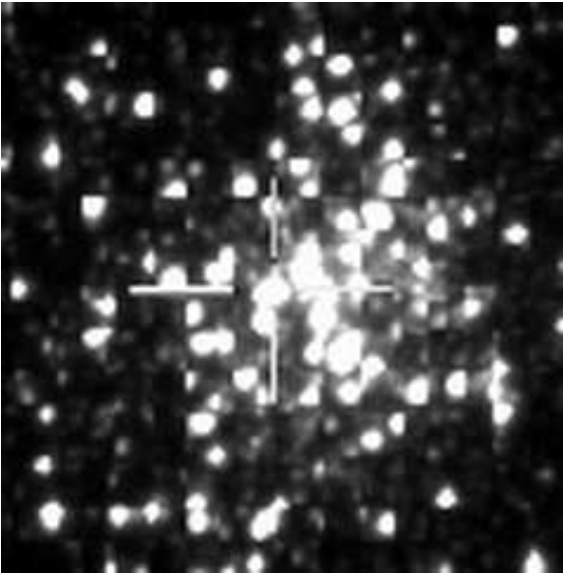}}
\caption{ Finder Chart for V4,
deep inside the NGC 1850 region, 
in a 60 arcsec square field, with North up
and East left.  The contrast has been adjusted so that the 
bright stars of the core region of NGC 1850 can be seen well here.  }
\label{mar04ebs-find}
\end{figure}

\begin{figure}
  \epsscale{.450}
  \rotatebox{90}{\plotone{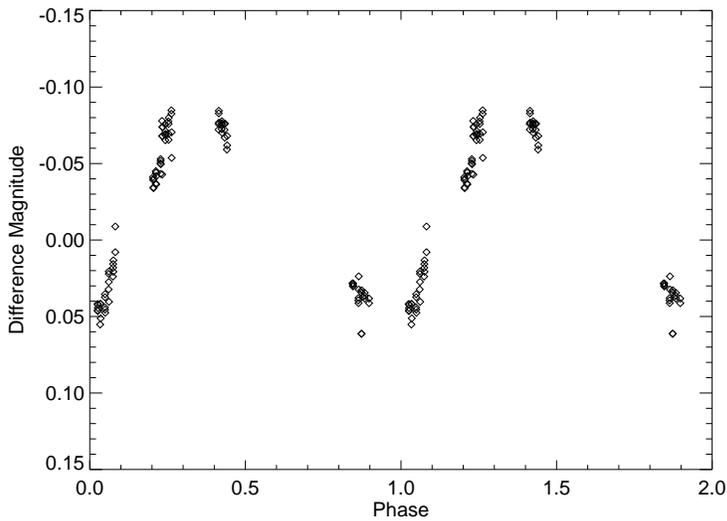}}
  \caption{ $I$-band difference magnitude phased light curve of V4, a 
    Cepheid variable also in the OGLE Cepheid Variables online catalog
    (Udalski et al. 1999). 
  }
  \label{stv004-dmag}
\end{figure}

V4 is another Cepheid variable discovery of OGLE, having a period 
of 5.57 days \citep{uda99}, and is shown in 
Figure \ref{mar04ebs-find}, which has had the
contrast adjusted to more easily see the bright stars such as V4.  
The light curve of V4 is shown in Figure \ref{stv004-dmag}.

\section{The LMC Field Eclipsing Binary}

\begin{figure}
\epsscale{.50}
\resizebox*{0.45\textwidth}{!}
    {\includegraphics{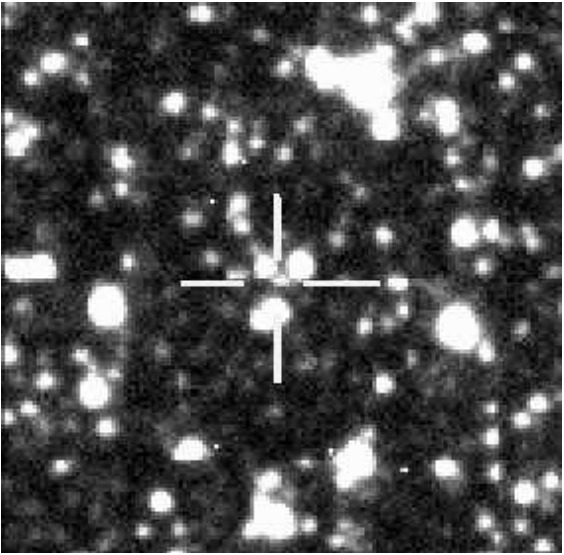}}
\caption{ Finder chart for V3, the detached eclipsing binary found among LMC field stars in a 60 arcsec square field, with North up
and East left. }
\label{stv003-find}
\end{figure}

\begin{figure}
  \epsscale{.450}
  \rotatebox{90}{\plotone{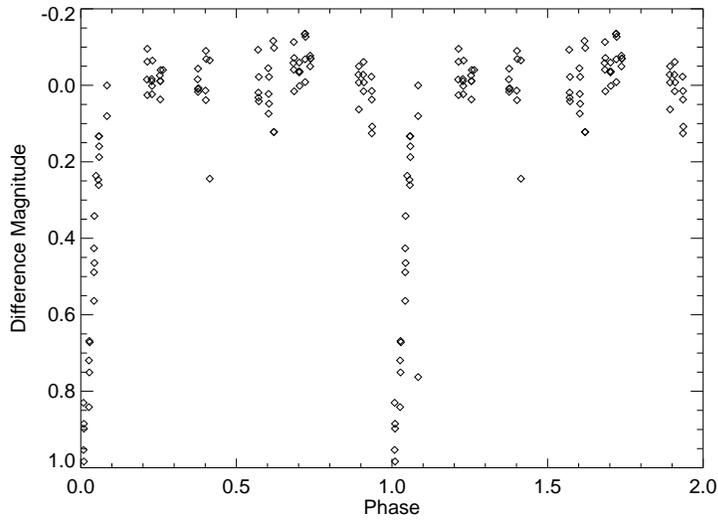}}
  \caption{ $I$-band difference magnitude phased light curve of V3, an 
    LMC field detached eclisping binary with dramatic dimming.  
    The primary luminosity dip is seen clearly in this data from 
    image-subtracted frames, despite V3 having a 
    PSF that is badly blended with brighter neighboring
    stars.
  }
  \label{stv003-dmag}
\end{figure}

The light curve of an LMC field eclipsing binary star, V3 
(finder chart Figure \ref{stv003-find}), 
is shown in Figure \ref{stv003-dmag}. 
It is presented as an excellent example
of an eclipsing binary which shows up clearly in image subtracted frames.
The MACHO project, using traditional photometry, has V3
catalogued on their website but does 
not identify it as an eclipsing binary.
The OGLE project, using image subtraction, was able to identify
V3 as an eclipsing binary even though it is on the edge of their
frame.  The power of image subtraction became apparent when in the
very early stages of the image subtraction analysis, this star
stood out strongly despite its being significantly fainter than 
the two or three nearby stars with significant blending; at that
time the OGLE finding was not known to the author
but the lower level of brightness
was apparent to the eye on the MACHO website's light curve of V3, and 
the stars dimming is apparent in the original images.
The current light curve produced the same 3.11 day period
and greater than 1 magnitude change in brightness, but does not cover the
secondary minimum.

\section{Motivation: Globular Cluster Evolution}


Study of MC GLCs is needed to answer the question of whether they 
represent a younger stage of GC evolution or if they are really dense
open clusters unrelated to true GCs.  Study of whether young GLCs,
intermediate age GLCs, and ``old'' Milky Way GCs have related binary 
fractions could help answer this question.  Comparisons of 
observations with models of the binary fraction of GC
evolution such as studied by \citet{fre03} are needed.  This would include 
both CMD analysis such as \citet{els98} and the somewhat more direct 
method of finding binary fractions from measurements of the 
fraction of stars that are eclipsing binaries.


Previously surveyed GCs for photometric 
binaries include 47 Tuc 
(Albrow et al. 2001; Kaluzny et al. 1997a, 1998; 
Edmonds et al. 1996), NGC 6934 (Kaluzny, Olech, \& Stanek 2001), 
M22 (Kaluzny and Thompson 2001), 
$\omega$ Centauri (Kaluzny et al. 1996, 1997b), 
M5 (Yan \& Reid 1996), M71 (Yan \& Mateo 1994), 
NGC 6397 (Rubenstein \& Bailyn 1996; Kaluzny 1997), 
M4 \citep{kal97,fer04}, NGC 3201 \citep{von02,von03}, 
NGC 4372 (Kaluzny \& Krzeminsky 1993), M10 \citep{von03}, 
and M12 \citep{von03}, with a compilation of 
W UMa-type (contact or semi-detached) binary stars compiled and 
discussed by Rucinski (2000).  GC binary fractions have also 
been studied using CHANDRA in the X-Ray region, 
including by \citet{poo03}.  
Most of the photometric work was done by the 
conventional technique of first determining the numerical 
magnitudes and then looking for periods in the photometric 
time series.  However, 
Kaluzny, Olech, \& Stanek (2001) and Albrow et al. (2001) 
were able to use Alard \& Lupton's 
improved method of image subtraction
\citep{ala98,ala99,ala00}.  They started 
their analysis with conventional techniques, but after using 
image subtraction were able to find 6 additional variable stars 
not found with their previous procedures. The work by 
Kaluzny, Olech, \& Stanek is of particular interest because 
it was done with a modest sized ground-based telescope (1.2 m).
Other authors who have improved their
results using image subtraction include 
\citet{woz00,moc01a,moc01b,kal01a,zeb01,kal01b}; and \citet{bru03}.
Because traditional crowded-field photometry (PSF or aperture 
fitting) works poorest in crowded fields, finding photometric
binaries in dense GCs is well suited for 
image-subtraction.
Comparative studies of variable stars in young GLCs are lacking 
compared to the studies of Milky Way GCs such as listed above.

Observations with both moderate and 
larger telescopes can then be used to evaluate CMD (Color Magnitude Diagram) 
studies of young GCs, 
and results from both can be compared with results 
from the relatively better studied ``old'' GCs of the Milky Way. 


%
%

\section{Future Work: Binary Fraction}

The current work shows that the binary fraction of a GLC
much younger than Milky Way GCs can be directly addressed with current
techniques, in addition to CMD analysis.  
Because of their importance to the study of GC evolution, 
longer duty-cycle surveys of MC GLCs are needed to determine 
the binary fraction of stars in 
NGC 1850 and similarly dense MC clusters in order to understand 
whether these binary fractions are consistent with 
dynamical models of GC evolution, and whether MC GLC and MW GC 
binary fractions as a function of age follow a consistent pattern.
Spectroscopic study of V5 and V1 would give important data 
about the distance to the well studied GLC NGC 1850.
These observations demonstrate than an extension of binary fraction
measurements to young clusters, even in regions of crowded stellar
fields, is now possible.

\acknowledgments

I am grateful to Alberto Miranda and the staff at the 
National Science Foundation's Cerro Tololo Inter-American Observatory,
in particular Edgardo Cosgrove, Manuel Hernandez, and Malcolm Smith,
for their assistance.  The generous encouragement and 
support of research at
Georgia State University under Harold A. McAlister
is acknowledged.  The support for using the CTIO 0.9 m
telescope by Todd Henry 
and the SMARTS Consortium is also
gratefully acknowledged.  CHARA members Doug Gies and William
Bagnuolo provided helpful discussion.  Computer network support by
John McFarland, Rajesh Deo,
James P. Kinney III, and Ginny Mauldin-Kinney is appreciated.
This research was supported by a 
Research Program Enhancement grant
from the Georgia State University Research Office.
This paper utilizes work from the OGLE project 
(Udalski et al. 1999; Udalski, 2003; Wyrzykowski et al. 2003). 
This paper utilizes public domain data originally obtained by the MACHO Project, whose work was performed under the joint auspices of the U.S. Department of Energy, National Nuclear Security Administration by the University of California, Lawrence Livermore National Laboratory under contract No. W-7405-Eng-48, the National Science Foundation through the Center for Particle Astrophysics of the University of California under cooperative agreement AST-8809616, and the Mount Stromlo and Siding Spring Observatory, part of the Australian National University.  
I appreciate the contributions of analysis code, OGLE data, and 
valuable suggestions from
Kaspar von Braun, Michael Reid, Hans Bruntt, Barbara J. Mochejska, 
Christophe Alard, and Andrzej Udalski.

\begin{table}[t]
\caption{ Number of Observations 
}
\vspace*{2mm}
{\centering 
\begin{tabular}{ccc}
\tableline \tableline
   Filter  & Exposure Time$^a$ & Number of Frames \\
   \tableline
   $I_C$     & 100  	   & 99  \\
   $I_C$     & 300  	   & 73  \\
\tableline \tableline
  \end{tabular}
	\newline \indent $^a$ Seconds
\par}
  \label{obsTable}
\end{table}

\begin{table}[t]
\caption{ Variables' Properties (Preliminary OGLE III data) 
}
\vspace*{2mm}
{\centering 
\begin{tabular}{ccccc}
\tableline \tableline
 Variable& Variable Type& Period (d) & Amplitude(s) ($I$)    \\
   \tableline
   1     & Detached EB  & 1.48       & 0.7, 0.6              \\
   2     & Cepheid      & 8.56       & 0.5                   \\
   3     & Detached EB  & 3.11       & 1.7, 1.1              \\
   4     & Cepheid      & 5.57       & 0.2                  \\
   5     & Detached EB  & 3.13	      & $0.35, 0.25         $  \\
\tableline \tableline
  \end{tabular}
\par}
  \label{varsTable}
\end{table}

\begin{table}[t]
\caption{ Variables' OGLE II Photometry
}
\vspace*{2mm}
{\centering 
\begin{tabular}{cccccc}
\tableline \tableline
 Variable& $V$         & $B$    & $I$     &  $B-V$ \\
   \tableline
   1     & $18.09 \pm 0.17$ & $18.16 \pm 0.20$ & $18.16 \pm 0.13$ & $+0.07$  \\
   2     & $14.65 \pm 0.36$ & $15.34 \pm 0.24$ & $13.91 \pm 0.15$ & $+0.70$  \\
   3     & ...           & ...           & $18.02 \pm 0.65 $ & ...     \\
   4     & $14.49 \pm 0.11$ & $14.97 \pm 0.09$ & $13.89 \pm 0.07$ &$+0.47$ \\
   5     & $14.40 \pm 0.12$ & $14.23 \pm 0.10$ & $14.55 \pm 0.08$ & $-0.17$ \\
\tableline \tableline
  \end{tabular}
\par}
  \label{oglePhotTable}
\end{table}

\begin{table}[t]
\caption{ OGLE, MACHO, and Previous Identifiers of Variables
}
\vspace*{2mm}
{\centering 
\begin{tabular}{cccc}
\tableline \tableline
 Variable& ID      & MACHO ID            & OGLE ID \\
   \tableline
   1     & ...     & ...                 & OGLE050842.01-684456.1\\
   2     & SW 58   & MACHO 1.4540.17     & LMC118.2 18259\\
   3     & ...     & ...                 & LMC110.7 1437\\       
   4     & ...     & ...                 & OGLE050846.31-684539.7\\
   5     & ...     & ...                 & LMC\_SC11 162262\\
\tableline \tableline
  \end{tabular}
\par}
  \label{varsIDtable}
\end{table}

\begin{table}[t]
\caption{ Positions of Variables, 2000
}
\vspace*{2mm}
{\centering 
\begin{tabular}{ccc}
\tableline \tableline
   Variable  & R.A.  Decl.  \\
   \tableline
   1         & 05 08 42.01 -68 44 56.1  	      \\
   2         & 05 08 43.17 -68 45 33.2  	      \\
   3         & 05 09 59.80 -68 44 25.3  	      \\
   4         & 05 08 46.31 -68 45 39.8  	      \\
   5         & 05 08 38.94 -68 45 45.7  	      \\
\tableline \tableline
  \end{tabular}
\par}
  \label{positionsTable}
\end{table}

\clearpage

\end{document}